\documentstyle[preprint,prb,aps]{revtex}
\tightenlines
\begin{document}
\draft
\title{All-electron magnetic response with 
pseudopotentials: NMR chemical shifts}
\author{Chris J. Pickard\cite{p-address}} 
\address{Institut f\"ur Geowissenschaften,\\ 
Universit\"at Kiel, Olshausenstrasse 40 D-24098 Kiel, Germany}
\author{Francesco Mauri} 
\address{Laboratoire de Min\'{e}ralogie-Cristallographie de Paris,
\\ Universit\'{e} Pierre et Marie Curie, 4 Place Jussieu, 75252, 
Paris, Cedex 05, France}
\date{\today}
\maketitle
\begin{abstract}
A theory for the \emph{ab initio} calculation of all-electron NMR
chemical shifts in insulators using pseudopotentials is presented. It
is formulated for both finite and infinitely periodic systems and is
based on an extension to the Projector Augmented Wave approach of
Bl\"{o}chl [P. E. Bl\"{o}chl, Phys. Rev. B {\bf 50}, 17953 (1994)] and
the method of Mauri \emph{et al} [F. Mauri, B.~G. Pfrommer, and
S.~G. Louie, Phys. Rev. Lett. {\bf 77}, 5300 (1996)]. The theory is
successfully validated for molecules by comparison with a selection of
quantum chemical results, and in periodic systems by comparison with
plane-wave all-electron results for diamond.
\end{abstract}

\newcommand{\ket}    [1]{{|#1\rangle}}
\newcommand{\bra}    [1]{{\langle#1|}}
\newcommand{\braket} [2]{{\langle#1|#2\rangle}}
\newcommand{\bracket}[3]{{\langle#1|#2|#3\rangle}}
\narrowtext

\section{Introduction}
\label{intro}

The experimental technique of nuclear magnetic resonance (NMR) is
widely used in structural chemistry and increasingly in solid state
studies.\cite{encyclopedia} Chemical shift ($\sigma$) spectra give
information about the atomic structure of the sample under
investigation. In the case of molecular systems, empirical rules are
commonly used to extract this information from the raw experimental
data. However, this approach cannot be applied in the solid state, as
the atomic configurations often cannot be modeled by chemical
analogues or reference compounds. In these cases \emph{ab initio}
calculations of the chemical shifts are the only way to obtain an
unambiguous determination of the microscopic structure.

Until recently, there has been no theory for the calculation of NMR
chemical shifts in extended periodic systems, and the conventional
approach to the theoretical interpretation of solid state NMR spectra
has been to approximate the infinite solid by a
cluster.\cite{nmr:tossell2} In this way, the traditional quantum
chemical approaches \cite{kutzelnigg90,giao,igaim,igaimgaussian94} can
be used to calculate the chemical shifts. Unfortunately, true
convergence with respect to basis set and cluster size is often not
possible due to the limitations of available computational resources.

The work of Mauri, Pfrommer and Louie\cite{mauri96II} solved the
problem of calculating NMR chemical shifts in the solid state with an
all-electron Hamiltonian.  Integrated with their approach to the
calculation of magnetic susceptibility,\cite{mauri96I} they presented
a theory for the \emph{ab initio} computation of NMR chemical shifts
in condensed matter systems using periodic boundary conditions
(hereafter referred to as the MPL method).  Although the MPL theory
has been derived using an all-electron Hamiltonian, so far, it has
only been implemented in an electronic structure code based on
norm-conserving pseudopotentials.  In such implementation the
complications inherent within the pseudopotential approximation have
been neglected.  For this reason, while several useful applications
have emerged, \cite{mauri97,yoon98,mauri99,water,rodo,mauri2000} the
method's use has been restricted to the calculation of chemical shifts
of light elements (hydrogen, carbon, and nitrogen) and of
silicon. Moreover, the description of the silicon chemical shifts
required the explicit inclusion of the 2$s$ and 2$p$ silicon orbitals
as valence and the use of a very high, and computationally expensive,
plane-wave energy-cutoff of 600 Ry.\cite{mauri2000} In the
above applications of the MPL method the pseudopotential error had
been assumed to be small and controllable.  To compute the NMR
chemical shifts of nuclei heavier than neon and to truly exploit the
ability of pseudopotentials to calculate the properties of complex,
low symmetry structures (which is well established for a wide range of
structural properties), a theory is required which does not ignore the
pseudopotential approximation.

Apart from the early and isolated attempt of Ridard, Levy and
Millie,\cite{pseudoridard} it has been widely expected within the
quantum chemical community that any theory for the calculation of NMR
chemical shifts for nuclei described with a
pseudopotential\cite{pseudomalkin} would fail due to the non-rigid
nature of the core contributions to the total chemical
shift.\cite{kutzelnigg90} However, a careful separation of core and
valence contributions that ensures that they are individually
gauge-invariant, by Gregor, Mauri and Car,\cite{gregor99} has shown
that this is not the case and that the core contributions are
rigid. This suggests that a pseudopotential based theory of NMR might,
in fact, exist.

One of the most obvious deficiencies of the pseudopotential approach
is that the pseudopotential approximation explicitly neglects the form
of the electronic wavefunctions near the nucleus. The pseudo
wavefunctions are chosen to be as smooth as possible in the core
region, and the correct nodal structure of the wavefunctions is
lost. This leads to a good approximation for the calculation of total
energies and their derivatives, and properties for which the matrix
elements are dominated by the regions outside the core. However, the
quantitative calculation of many properties --- hyperfine parameters,
core level spectra, electric-field-gradients and the NMR chemical
shifts --- depend critically on the details of the all-electron
wavefunctions at the nucleus. Van de Walle and Bl\"{o}chl presented a
solution to this problem for the calculation of hyperfine
parameters\cite{vandewalle93} based on Bl\"{o}chl's Projector
Augmented-Wave (PAW) electronic structure method,\cite{blochl94} which
is itself closely related to Vanderbilt's ultrasoft pseudopotential
scheme.\cite{vanderbilt90} While in all but a few reported cases,
where core-electron polarization effects are
important\cite{holzwarth97} or in some magnetic
systems,\cite{kresse99} the PAW method gives similar results to
pseudopotential approaches, it does provide a extremely useful
framework for the unification of all-electron (Full-potential)
Linearized Augmented Plane-Wave\cite{blaha90} and pseudopotential
approaches. Indeed, it it becoming clear that the PAW approach, which
will be described in more detail in Section \ref{paw-method}, offers a
general approach to the calculation of all-electron properties from
pseudopotential based schemes. Following the work of Van de Walle and
Bl\"{o}chl, core level spectra,\cite{phd:pickard,pickard97} momentum
matrix elements,\cite{kageshima97} and electric field
gradients\cite{petrilli98} have all been calculated using the PAW
scheme.

In this paper we present a theory for all-electron magnetic response
within the pseudopotential approximation and its application to the
calculation of first principles NMR chemical shifts. The connection
between the current response and the chemical shifts is outlined in
Section \ref{nmrshifts}. We introduce an extension of Bl\"{o}chl's PAW
approach, which we call the Gauge Including Projector Augmented-Wave
(GIPAW) approach. This will be described in Section \ref{pspinb}. A
Hamiltonian constructed using GIPAW has the required translational
invariance in the presence of a magnetic field. This is not true for
the original PAW formulation. In Section \ref{current-finite} we
present our theory for finite systems. In Section \ref{current-extend}
we reformulate our expressions for extended systems. To be useful,
these expressions must be restricted to \emph{periodic} extended
systems, and the periodic theory is presented in Section
\ref{current-pbc-extend}. Both the theories for finite and extended
periodic systems summarized in Section \ref{summary} have been
implemented in a plane-wave pseudopotential electronic structure
code. Details of our implementation are given in Section
\ref{calc-nmrshifts}. We validate the method by comparison with
calculations by Gregor \emph{et al}\cite{gregor99} for a selection of
small molecules. The theory for extended systems is further validated
by comparison to results obtained by an all-electron plane-wave
calculation for a crystalline material, diamond.

\section{NMR chemical shifts}
\label{nmrshifts}

A uniform, external magnetic field ${\bf B}$ applied to a sample of
matter induces an electric current. In an insulating non-magnetic
material, only the orbital motion of the electrons contribute to this
current.  Moreover, for the field strengths typically used in NMR
experiments, the induced electronic current is proportional to the
external field ${\bf B}$ and is the first order induced current, ${\bf
j}^{(1)}({\bf r}')$.  The current ${\bf j}^{(1)}({\bf r}')$ produces a
non-uniform magnetic field,
\begin{equation}
{\bf B}^{(1)}_{\rm in}({\bf r})=\frac{1}{c}\int d^3r' {\bf
j}^{(1)}({\bf r}')\times \frac{{\bf r}-{\bf r}'}{|{\bf r}-{\bf r}'|^3}.
\label{bs}
\end{equation}
The chemical shift is defined as the ratio between the induced
magnetic field and the external uniform applied magnetic field:
\begin{equation}
{\bf B}^{(1)}_{\rm in}({\bf r})=-{\tensor\sigma}({\bf r}){\bf B}.
\label{bs2}
\end{equation}
Here ${\tensor \sigma}({\bf r})$ is the chemical shift tensor, and the
isotropic chemical shift is given by $\sigma({\bf r})={\rm Tr} [{\tensor
\sigma}({\bf r})]/3$.  NMR experiments can measure $\tensor
\sigma({\bf r})$ at the nuclear positions.  To compute the chemical
shift tensor we first obtain ${\bf j}^{(1)}({\bf r})$ by perturbation
theory and then we evaluate ${\bf B}^{(1)}_{\rm in}({\bf r})$ using
Eq. (\ref{bs}). We now describe our new approach to the calculation of
an induced all-electron current ${\bf j}^{(1)}({\bf r}')$ using
pseudopotentials and in Section \ref{calc-nmrshifts} the computational
procedure we use to obtain ${\bf j}^{(1)}({\bf r}')$, and finally
$\sigma({\bf r})$, is detailed.

\section{Pseudopotentials in a magnetic field}
\label{pspinb}

In this section we develop the Gauge Including Plane-Wave method,
first describing the original Projector Augmented-Wave method, and
then extending it to the case of a uniform applied magnetic field.

\subsection{Projector augmented-wave method}
\label{paw-method}

In Ref.~\onlinecite{blochl94}, Bl\"ochl introduced a linear
transformation operator ${\cal T}$ which maps the valence pseudo
wavefunctions $|\tilde \Psi\rangle$ onto the corresponding
all-electron wavefunctions, $|\Psi\rangle={\cal T}|\tilde
\Psi\rangle$.  The operator is defined by specifying a set of target
all-electron partial waves $|\phi_{{\bf R},n}\rangle$ obtained by the
application of ${\cal T}$ on to a set of pseudo partial waves
$|\tilde\phi_{{\bf R},n}\rangle$ with
\begin{equation}
{\cal T}={\bf 1}+\sum_{{\bf R},n} [|\phi_{{\bf R},n}\rangle -
|\tilde\phi_{{\bf R},n}\rangle]\langle\tilde p_{{\bf R},n}|
\label{paw}
\end{equation}
and $\langle\tilde p_{{\bf R},n}|$ are a set of projectors such that
$\langle\tilde p_{{\bf R},n}|\tilde\phi_{{\bf R}',m}\rangle=
\delta_{{\bf R},{\bf R}'} \delta_{n,m}$.  Each projector and partial
wave is an atomic-like function centered on an atomic site ${\bf R}$,
and the index $n$ refers to the angular momentum quantum numbers and
to an additional number used if there are more than one projector per
angular momentum channel. The expectation value of an operator $O$
between all-electron wavefunctions can be expressed as the expectation
value of a pseudo operator $\tilde O={\cal T}^+ O{\cal T}$ between the
corresponding pseudo wavefunctions.

To obtain a useful formalism we must make some further assumptions.
In particular, for each atomic site we define an augmentation region
$\Omega_{\bf R}$ and suppose that: i) outside the augmentation region
$\Omega_{\bf R}$, the $|\tilde\phi_{{\bf R},n}\rangle$ coincide with the
$|\phi_{{\bf R},n}\rangle$, ii) outside the augmentation region
$\Omega_{\bf R}$, the $|\tilde p_{{\bf R},n}\rangle$ vanish, iii) within
the augmentation region $\Omega_{\bf R}$, the $|\phi_{{\bf R},n}\rangle$
form a complete set for the valence wavefunctions, i.e. any physical
valence all-electron wavefunction can be written, within $\Omega_{\bf R}$,
as a linear combination of all-electron partial waves, and finally iv)
the augmentation regions of different sites do not overlap.  Bl\"ochl
has shown that given these assumptions, if $O$ is a local or a
semi-local (such as ${\bf p}$ or $p^2$) operator:
\begin{equation}
\tilde O= O+\sum_{{\bf R},n,m}|\tilde p_{{\bf R},n}\rangle[\langle
\phi_{{\bf R},n}|O|\phi_{{\bf R},m}\rangle -\langle \tilde \phi_{{\bf
R},n}|O|\tilde\phi_{{\bf R},m}\rangle]\langle \tilde p_{{\bf R},m}|.
\label{psoperator}
\end{equation}

For simplicity, we shall further suppose that the norms computed
within $\Omega_{\bf R}$ of $|\tilde \phi_{{\bf R},n}\rangle$ and
$|\phi_{{\bf R},n}\rangle$ coincide. We then recover the norm
conserving pseudopotential formalism in the Kleinman-Bylander
\cite{kleinbyl} form.  The pseudo wavefunctions which correspond to
the all-electron valence eigenstates of the all-electron Hamiltonian
$H$ are eigenstates of the pseudo Hamiltonian $\tilde H$ with the same
eigenvalues. In the absence of a magnetic field the pseudo Hamiltonian
is:
\begin{equation}
\tilde H={\cal T}^+ H{\cal T}=\frac{1}{2}{\bf p}^2 +V^{\rm loc}({\bf
r})+ \sum_{\bf R}V^{\rm nl}_{\bf R},
\label{pseudoHnoB}
\end{equation}
where ${\bf p}$ is the momentum operator, and $V^{\rm loc}({\bf r})$
is the local part of the pseudopotentials, which includes the
self-consistent part of the Hamiltonian. The non-local part of the
pseudopotential at the atomic site ${\bf R}$ in the above expression
is,
\begin{equation}
V^{\rm nl}_{\bf R}=\sum_{n,m} |\tilde p_{{\bf R},n}\rangle a^{\bf R}_{n,m}
\langle \tilde p_{{\bf R},m}|.
\label{nonlocpot}
\end{equation}
The $a^{\bf R}_{n,m}$ are the strengths of the non-local potential in
each channel, and the depend on ${\bf R}$ since each atomic site may
be occupied by a different chemical species.

The choice of the pseudo partial waves and projectors is largely
arbitrary. However, for a scheme to be useful, all the lowest
eigenvalues of $\tilde H$ should coincide with a valence eigenvalue of
$H$ up to an given energy $E_{\rm val}^{\rm max}$, i.e. no ghost states
should be introduced in the pseudo spectrum up to an energy $E_{\rm
val}^{\rm max}$.  The energy $E_{\rm val}^{\rm max}$ depends on the
specific property we wish to compute, and should at least be larger
than the highest occupied eigenvalue.

In contrast to the traditional formulation of pseudopotentials, using
the PAW formulation it is possible to obtain the expectation values of
all-electron operators in terms of pseudo wavefunctions using the
pseudo operators defined in Eq. (\ref{psoperator}).

\subsection{A single augmentation region in a uniform magnetic field}
\label{oneregion}

In presence of a uniform external magnetic field ${\bf B}$ the
all-electron Hamiltonian is:
\begin{equation}
H=\frac{1}{2}\left({\bf p}+\frac{1}{c}{\bf A}({\bf r})\right)^2+V({\bf
r}),
\end{equation}
where $c$ is the speed of light, $V({\bf r})$ is the all-electron
local potential, and ${\bf B}=\nabla \times {\bf A}({\bf r})$.  We
want to construct the corresponding pseudo Hamiltonian for a complex
system, which will contain many augmentation regions.  However, before
treating this general case, we consider a simplified system with just
a single augmentation region. The spatial origin is chosen to coincide
with the atomic site of the augmentation region.  In the symmetric
gauge ${\bf A}({\bf r})= \frac{1}{2}{\bf B}\times ({\bf r}-{\bf d})$,
where ${\bf d}$ is a constant vector which indicates the gauge origin.
The expectation values of the all-electron eigenstates for observable
operators do not depend on the gauge origin ${\bf d}$.  However, the
number of partial waves required to correctly describe the valence
all-electron eigenstates in the augmentation region critically depends
on the choice of ${\bf d}$.  To minimize the number of partial waves
required we must put the gauge origin at the atomic site of the
augmentation region, setting ${\bf d}={\bf 0}$.  Making this choice,
we minimize the effect of the magnetic field on the all-electron
wavefunctions in the augmentation region, where $|{\bf A}({\bf r})|^2$
and its spatial derivatives attain their minimum value. Moreover, with
this choice of gauge, the interaction between the valence and core
states of the augmented atom is negligibly small.\cite{gregor99} This
is essential if we are to make the pseudopotential approximation.
With
\begin{equation}
{\bf A}({\bf r})=\frac{1}{2}{\bf B}\times {\bf  r},
\label{A}
\end{equation}
the all-electron Hamiltonian becomes:
\begin{equation}
H=\frac{1}{2}{\bf p}^2 +V({\bf r})+\frac{1}{2c}{\bf L}\cdot { \bf
B}+\frac{1}{8c^2}({\bf B}\times {\bf r})^2,
\end{equation}
where $\bf L={\bf r}\times {\bf p}$ is the angular momentum operator
computed with respect to the atomic site within the augmentation
region. Using Eq. (\ref{psoperator}) and Eq. (\ref{pseudoHnoB}), we
obtain the corresponding pseudo Hamiltonian:
\begin{equation}
\tilde H=\frac{1}{2}{\bf p}^2 +V^{\rm loc}({\bf r})+ V^{\rm nl}_{\bf
0}+\frac{1}{2c}{\bf L}\cdot { \bf B}+\frac{1}{8c^2}({\bf B}\times {\bf
r})^2+\sum_{n,m} |\tilde p_{{\bf
0},n}\rangle(b^{(1)}_{n,m}+b^{(2)}_{n,m}) \langle \tilde p_{{\bf
0},m}|,
\end{equation}
where 
\begin{equation}
b^{(1)}_{n,m}=\frac{1}{2c} { \bf B} \cdot[\langle \phi_{{\bf
0},n}|{\bf L}|\phi_{{\bf 0},m}\rangle -\langle \tilde \phi_{{\bf
0},n}|{\bf L}|\tilde\phi_{{\bf 0},m}\rangle]
\end{equation}
and
\begin{equation}
b^{(2)}_{n,m}=\frac{1}{8c^2}[\langle \phi_{{\bf 0},n}|({\bf B}\times
{\bf r})^2|\phi_{{\bf 0},m}\rangle -\langle \tilde \phi_{{\bf
0},n}|({\bf B}\times {\bf r})^2|\tilde\phi_{{\bf 0},m}\rangle]
\end{equation}
If just one projector per angular momentum channel is used, as is
usually the case with norm conserving
pseudopotentials,\cite{hsc:vps,tm:vps} $b^{(1)}_{n,m}$ exactly
vanishes, since $|\phi_{{\bf 0},n}\rangle$ and $|\tilde\phi_{{\bf
0},n}\rangle$ are eigenstates of $L$ and $L_z$ with the same norm
within the augmentation region.  Moreover, since $({\bf B}\times {\bf
r})^2$ goes to zero in the center of the augmentation region, for norm
conserving pseudopotentials the term $b^{(2)}_{n,m}$ can also be
neglected.  Thus, with one augmentation region centered at the gauge
origin, the coupling with the magnetic field in the pseudo and
all-electron Hamiltonians has the same form, i.e.:
\begin{equation}
\tilde H=\frac{1}{2}{\bf p}^2 +V^{\rm loc}({\bf r})+ V^{\rm nl}_{\bf
0}+\frac{1}{2c}{\bf L}\cdot { \bf B}+\frac{1}{8c^2}({\bf B}\times {\bf
r})^2.
\label{Honereg}
\end{equation}

\subsection{Translations in a uniform magnetic field}
\label{translations}

The derivation in the previous section is not useful for systems with
several augmentation regions. Indeed, the gauge origin can coincide
with just one augmentation site at any given time. As a result, the
number for projectors of the other augmentation regions would have to
be increased to reach completeness in those regions. The cause of this
problem is that the PAW approach does not preserve translational
invariance in a uniform magnetic field.

In a uniform magnetic field the description of the system should be
invariant upon a rigid translation of all the atoms by a vector ${\bf
t}$.  Following the translation, the all-electron potential becomes
$V'({\bf r})=V({\bf r}-{\bf t})$ and the corresponding Hamiltonian is:
\begin{equation}
H'=\frac{1}{2}\left({\bf p}+\frac{1}{c}{\bf A}({\bf
r})\right)^2+V({\bf r}-{\bf t}),
\end{equation}
where ${\bf A}({\bf r})$ is still given by Eq. (\ref{A}).  Because of
the translational invariance, the eigenenergies of $H'$ coincide with
the eigenenergies of the original Hamiltonian $H$.  However, the new
eigenstates $|\Psi_n'\rangle$, are not just obtained by a rigid
translation of the original eigenstates $|\Psi_n\rangle$, but, upon
translation, they pick up an additional phase factor proportional to
the magnetic field:
\begin{equation}
\langle {\bf r}|\Psi_n'\rangle=e^{\frac{i}{2c}{\bf r}\cdot{\bf
t}\times{\bf B}}\langle {\bf r}-{\bf t}|\Psi_n\rangle.
\label{trans}
\end{equation}
The PAW transformation does not ensure exact invariance upon
translation, since the pseudo wavefunctions constructed with the
$\cal T$ transformation operator of
Eq. (\ref{paw}) do not transform according to Eq. (\ref{trans}).

\subsection{Gauge including projector augmented-wave method}
\label{gipawmethod}

To restore the translational invariance within a PAW-like approach, we
introduce a field dependent transformation operator ${\cal T}_{{\bf
B}}$, which, by construction, imposes the translational invariance
exactly:
\begin{equation}
{\cal T}_{{\bf B}}={\bf 1}+\sum_{{\bf R},n} e^{\frac{i}{2c}{\bf
r}\cdot{\bf R}\times{\bf B}}[|\phi_{{\bf R},n}\rangle - |\tilde
\phi_{{\bf R},n}\rangle]\langle\tilde p_{{\bf
R},n}|e^{-\frac{i}{2c}{\bf r}\cdot{\bf R}\times{\bf B}}.
\label{gipaw}
\end{equation}
This new transformation defines our novel approach, which we call the
Gauge Including Projected Augmented-Wave (GIPAW) method.  
In the following, we indicate
with a bar the pseudo wavefunctions and operators obtained using
${\cal T}_{{\bf B}}$ operator by analogy to Bl\"{o}chl's use of the
tilde.  By construction, the pseudo eigenstates, $|\bar \Psi\rangle$,
generated from the all-electron eigenstates using $|\Psi\rangle={\cal
T}_{{\bf B}}|\bar \Psi\rangle$, satisfy the same translation relation
as the all-electron eigenstates given by Eq. (\ref{trans}).  The GIPAW
pseudo operator $\bar O={\cal T}_{{\bf B}}^+ O {\cal T}_{{\bf B}}$
corresponding to a local or a semi-local operator $O$ is given by:
\begin{eqnarray}
\bar O= O +\sum_{{\bf R},n,m}e^{\frac{i}{2c}{\bf r}\cdot{\bf
R}\times{\bf B}}|\tilde p_{{\bf R},n}\rangle
&&
\left[\langle \phi_{{\bf R},n}|e^{-\frac{i}{2c}{\bf r}\cdot{\bf
R}\times{\bf B}}Oe^{\frac{i}{2c}{\bf r}\cdot{\bf R}\times{\bf B}}
|\phi_{{\bf R},m}\rangle -\langle \tilde \phi_{{\bf
R},i}|e^{-\frac{i}{2c}{\bf r}\cdot{\bf R}\times{\bf
B}}Oe^{\frac{i}{2c}{\bf r}\cdot{\bf R}\times{\bf B}}|\tilde\phi_{{\bf
R},m}\rangle\right] \nonumber \\
&&  
\langle \tilde p_{{\bf R},m}|e^{-\frac{i}{2c}{\bf r}\cdot{\bf
R}\times{\bf B}}
\label{gipawpsoperator}
\end{eqnarray}
There are connections between our GIPAW approach, and the
gauge-including atomic orbitals\cite{giao} (GIAO) and the independent
gauge for localized orbitals\cite{kutzelnigg90} (IGLO) methods, widely
used in the quantum chemical community.  However, it should be
recognized that in GIPAW the phase required to maintain the
translational invariance is carried by the operators, whereas in the
GIAO and in IGLO approaches the field dependent phase is attached to
the basis functions and to the occupied electronic orbitals,
respectively.

\subsection{GIPAW Hamiltonian}
\label{gipawhamiltonian}

Using Eq. (\ref{gipawpsoperator}), the identity
\begin{equation}
e^{-\frac{i}{2c}{\bf r}\cdot{\bf R}\times{\bf B}}\left({\bf
p}+\frac{1}{c}{\bf A}({\bf r})\right)^ne^{\frac{i}{2c}{\bf r}\cdot{\bf
R}\times{\bf B}}=\left({\bf p}+\frac{1}{c}{\bf A}({\bf r}-{\bf
R})\right)^n,
\label{identity}
\end{equation}
for integer $n$, and the outcomes of the discussion concerning
$b^{(1)}_{n,m}$ and $b^{(2)}_{n,m}$ in Section \ref{oneregion}, we
finally obtain the GIPAW pseudo Hamiltonian:
\begin{equation}
\bar H=\frac{1}{2}{\bf p}^2 +V^{\rm loc}({\bf r})+ \sum_{\bf
R}e^{\frac{i}{2c}{\bf r}\cdot{\bf R}\times{\bf B}} V^{\rm nl}_{\bf
R}e^{-\frac{i}{2c}{\bf r}\cdot{\bf R}\times{\bf B}}+\frac{1}{2c}{\bf
L}\cdot { \bf B} +\frac{1}{8c^2}({\bf B}\times {\bf r})^2.
\label{gipawH}
\end{equation}

The GIPAW Hamiltonian coincides with the PAW Hamiltonian,
Eq. (\ref{pseudoHnoB}) for ${\bf B}=0$, and with the PAW Hamiltonian,
Eq. (\ref{Honereg}), for ${\bf B}\ne0$ in systems with a single
augmentation region centered at the origin.  Moreover, as expected,
the GIPAW eigenenergies are exactly invariant upon translation, in
contrast to the PAW eigenenergies.

For later use in perturbation theory, $\bar H$ can be expanded in
powers of $B$:
\begin{equation}
\bar H=\bar H^{(0)}+\bar H^{(1)} + O(B^2)
\end{equation}
where $\bar H^{(0)}=\tilde H^{(0)}$ is the unperturbed Hamiltonian
given by Eq. (\ref{pseudoHnoB}), and
\begin{equation}
\bar H^{(1)}= \frac{1}{2c}\left({\bf L}+ \sum_{\bf R}{\bf R} \times
{\bf v}^{\rm nl}_{\bf R}\right) \cdot { \bf B},
\end{equation}
where
\begin{eqnarray} 
{\bf v}^{\rm nl}_{\bf R}&=&\frac{1}{i}[{\bf r},V^{\rm nl}_{\bf R}],
\end{eqnarray} 
and with square brackets we indicate the commutator. 

\subsection{GIPAW Current operator}
\label{gipawcurrent}

Another observable required to compute the NMR chemical shifts is the
current.  The all-electron electric current operator evaluated at the
position ${\bf r}'$ is:
\begin{equation} 
{\bf J}({\bf r}')= {\bf J}^{\rm p}({\bf r}')-\frac{{\bf A}({\bf
r}')}{c}|{\bf r}'\rangle\langle{\bf r}'| ={\bf J}^{\rm p}({\bf r}')
-\frac{{\bf B}\times{\bf r}'}{2c}|{\bf r}'\rangle\langle{\bf r}'|,
\end{equation}
where ${\bf J}^{\rm p}({\bf r}')$ is the paramagnetic current operator,
\begin{equation}
{\bf J}^{\rm p}({\bf r}')=-\frac{{\bf p}|{\bf r}'\rangle\langle{\bf
r}'|+|{\bf r}'\rangle\langle{\bf r}'|{\bf p}}{2}.
\end{equation}
Using Eq. (\ref{gipawpsoperator}), and Eq. (\ref{identity}), we
obtain the corresponding GIPAW operator:
\begin{equation}
\bar{\bf J}({\bf r}')={\bf J}^{\rm p}({\bf r}')-\frac{{\bf
B}\times{\bf r}'}{2c}|{\bf r}'\rangle\langle{\bf r}'|+\sum_{{\bf
R}}e^{\frac{i}{2c}{\bf r}\cdot{\bf R}\times{\bf B}}\left[\Delta{\bf
J}^{\rm p}_{\bf R}({\bf r}')+\Delta{\bf J}^{\rm d}_{\bf R}({\bf
r}')\right]e^{-\frac{i}{2c}{\bf r}\cdot{\bf R}\times{\bf B}},
\end{equation}
where
\begin{equation}
\Delta{\bf J}^{\rm p}_{\bf R}({\bf r}')=\sum_{n,m}|\tilde p_{{\bf
R},n}\rangle \left[\langle \phi_{{\bf R},n}| {\bf J}^{\rm p}({\bf
r}')|\phi_{{\bf R},m}\rangle - \langle \tilde \phi_{{\bf R},n}| {\bf
J}^{\rm p}({\bf r}') |\tilde\phi_{{\bf R},m}\rangle\right]
\langle\tilde p_{{\bf R},m}|
\end{equation}
is what we call the paramagnetic augmentation operator, and
\begin{equation}
\Delta{\bf J}^{\rm d}_{\bf R}({\bf r}')=-\frac{{\bf B}\times({\bf
r}'-{\bf R})}{2c}\sum_{n,m}|\tilde p_{{\bf R},n}\rangle\left[\langle
\phi_{{\bf R},n}|{\bf r}'\rangle\langle{\bf r}'|\phi_{{\bf
R},m}\rangle - \langle \tilde \phi_{{\bf R},n}|{\bf
r}'\rangle\langle{\bf r}'|\tilde\phi_{{\bf R},m}\rangle\right]\langle
\tilde p_{{\bf R},m}|,\label{defDeltad}
\end{equation}
is  what we call the diamagnetic augmentation operator.

As for the Hamiltonian, for perturbation theory purposes it is
useful to expand the operator $\bar {\bf J}({\bf r})$ in powers of
$B$:
\begin{equation}
\bar {\bf J}({\bf r}')=\bar {\bf J}^{(0)}({\bf r}')+\bar {\bf
J}^{(1)}({\bf r}')+O(B^2),
\end{equation}
with
\begin{equation}
\bar {\bf J}^{(0)}({\bf r}')={\bf J}^{\rm p}({\bf r}')+\sum_{\bf R}
\Delta{\bf J}^{\rm p}_{\bf R}({\bf r}'),
\end{equation}
and
\begin{equation}
\bar {\bf J}^{(1)}({\bf r}')=-\frac{{\bf B}\times{\bf r}'}{2c}|{\bf
r}'\rangle\langle{\bf r}'|+\sum_{\bf R} \left[\Delta{\bf J}^{\rm
d}_{\bf R}({\bf r}')+\frac{1}{2ci}[{\bf B}\times {\bf R}\cdot{\bf
r},\Delta{\bf J}^{\rm p}_{\bf R}({\bf r}')]\right].
\end{equation}

\section{Current response in finite systems}
\label{current-finite}

Within density functional perturbation theory, the current can be
computed using the GIPAW operators and wavefunctions as:
\begin{equation}
{\bf j}^{(1)}({\bf r}')= 2\sum_o\left[ \langle \bar
\Psi_{o}^{(1)}|\bar{\bf J}^{(0)}({\bf r}')|\bar
\Psi_{o}^{(0)}\rangle+\langle \bar \Psi_{o}^{(0)}|\bar{\bf
J}^{(0)}({\bf r}')|\bar \Psi_{o}^{(1)}\rangle+\langle \bar
\Psi_{o}^{(0)}|\bar{\bf J}^{(1)}({\bf r}')|\bar
\Psi_{o}^{(0)}\rangle\right].
\label{perstart}
\end{equation}
Here the factor of two accounts for spin degeneracy and the sum runs
over the occupied orbitals $o$. The wavefunction
$|\bar\Psi_{n}^{(0)}\rangle$ is an unperturbed eigenstate of $\tilde
H^{(0)}$ with eigenvalue $\varepsilon_n$ and
$|\bar\Psi_{n}^{(1)}\rangle$ is its linear variation, projected in the
empty subspace:
\begin{equation}
|\bar \Psi_{n}^{(1)}\rangle = {\cal G}(\varepsilon_n)\bar H^{(1)}|
 \bar \Psi_{n}^{(0)}\rangle.
\end{equation}
The Green function operator is:
\begin{equation}
{\cal G}(\varepsilon)=\sum_{e}\frac{|\bar\Psi_{e}^{(0)}\rangle \langle
\bar \Psi_{e}^{(0)}|}{ \varepsilon-\varepsilon_e },
\label{greenfn}
\end{equation}
with the sum running over the empty orbitals $e$. Reordering the
different contributions of Eq. (\ref{perstart}) we obtain:
\begin{equation}
{\bf j}^{(1)}({\bf r}')={\bf j}^{(1)}_{\rm bare}({\bf r}')+{\bf
j}^{(1)}_{\rm \Delta p}({\bf r}')+{\bf j}^{(1)}_{\rm \Delta d}({\bf
r}'),
\end{equation}
where
\begin{equation}
{\bf j}^{(1)}_{\rm bare}({\bf r}')=4\sum_{o}{\rm Re}\left[\langle \bar
\Psi_{o}^{(0)}|{\bf J}^{\rm p}({\bf r}'){\cal G}(\varepsilon_o) \bar
H^{(1)}| \bar \Psi_{o}^{(0)}\rangle\right]-\frac{1}{2c}\rho^{\rm
ps}({\bf r}'){\bf B}\times {\bf r}'.
\label{jbare}
\end{equation}
$\rm Re$ stands for taking the real part and $\rho^{\rm ps}({\bf
r}')=2\sum_o \langle \bar \Psi_{o}^{(0)}|{\bf r}'\rangle\langle {\bf
r}' |\bar \Psi_{o}^{(0)}\rangle$ is the ground state pseudo
density. The paramagnetic correction to the current is
\begin{equation}
{\bf j}^{(1)}_{\rm \Delta p}({\bf r}')=\sum_{{\bf R'},o}\left\{4{\rm
Re} \left[\langle \bar \Psi_{o}^{(0)}|\Delta{\bf J}^{\rm p}_{\bf
R'}({\bf r}'){\cal G}(\varepsilon_o) \bar H^{(1)}| \bar
\Psi_{o}^{(0)}\rangle\right]+2\langle \bar
\Psi_{o}^{(0)}|\frac{1}{i2c}[{\bf B}\times {\bf R'}\cdot{\bf
r},\Delta{\bf J}^{\rm p}_{\bf R'}({\bf r}')]| \bar
\Psi_{o}^{(0)}\rangle\right\},
\label{jp}
\end{equation}
and the diamagnetic correction is
\begin{equation}
{\bf j}^{(1)}_{\rm \Delta d}({\bf r}')=2\sum_{{\bf R},o}\langle \bar
\Psi_{o}^{(0)}|\Delta{\bf J}^{\rm d}_{\bf R}({\bf r}')| \bar
\Psi_{o}^{(0)}\rangle.
\label{jd}
\end{equation}
Notice that the last two current contributions, ${\bf j}^{(1)}_{\rm
\Delta p}({\bf r}')$ and ${\bf j}^{(1)}_{\rm \Delta d}({\bf r}')$, are
written as a sum over augmentation sites and vanish outside the
augmentation regions, where the all-electron and pseudo partial waves
coincide.
 
By construction, the current ${\bf j}^{(1)}({\bf r}')$ computed within
the GIPAW formalism is, as all physical observables should be, invariant
upon translation of the system by a vector ${\bf t}$, i.e. after
translation the new current should be ${\bf j}^{(1)}({\bf r}'-{\bf
t})$. Interestingly, all three terms, ${\bf j}^{(1)}_{\rm bare}({\bf
r}')$, ${\bf j}^{(1)}_{\rm \Delta p}({\bf r}')$ , and ${\bf
j}^{(1)}_{\rm \Delta d}({\bf r}')$, are individually invariant upon
translation.  The invariance of ${\bf j}^{(1)}_{\rm \Delta d}({\bf
r}')$ is obvious from the definition of the $\Delta{\bf J}^{\rm d}_{\bf
R}({\bf r}')$ operator, Eq. (\ref{defDeltad}).  The invariance of the
other two contributions is less evident, and to prove it, we need to
manipulate Eqs. (\ref{jbare}) and (\ref{jp}).  To this end, we
notice that the second term in the r.h.s.  of Eq. (\ref{jbare}) can be
rewritten as a commutator, 
\begin{equation}
-\frac{1}{2c}\rho^{\rm ps}({\bf r}'){\bf B}\times {\bf r}'=2\sum_o
 \frac{1}{2c}\langle \bar \Psi_{o}^{(0)}|\frac{1}{i}[{\bf B}\times
 {\bf r}'\cdot{\bf r},{\bf J}^{\rm p}({\bf r}')]|\bar
 \Psi_{o}^{(0)}\rangle.
\end{equation}
We can now use the generalized $f$-sum rule established in Appendix
\ref{Asum-rule}, Eq. (\ref{f-sumrule}), with the operators ${\bf
J}^{\rm p}({\bf r}')$ and $\Delta{\bf J}^{\rm p}_{\bf R'}({\bf r}')$ in
the place of ${\cal O}$ and the operator ${\bf r}$ in the place of
${\cal E}$, to rewrite the second terms in the r.h.s. of both
Eqs. (\ref{jbare}) and (\ref{jp}), obtaining:
\begin{equation}
{\bf j}^{(1)}_{\rm bare}({\bf r}')=4\sum_{o}{\rm Re}\left[\langle \bar
\Psi_{o}^{(0)}|{\bf J}^{\rm p}({\bf r}'){\cal G}(\varepsilon_o) \bar
H^{(1)}| \bar \Psi_{o}^{(0)}\rangle-\langle \bar \Psi_{o}^{(0)}| {\bf
J}^{\rm p}({\bf r}'){\cal G}(\varepsilon_o) \frac{{\bf B}\times {\bf
r}'}{2c}\cdot {\bf v}| \bar \Psi_{o}^{(0)}\rangle\right],
\label{jbarebis}
\end{equation}
\begin{equation}
{\bf j}^{(1)}_{\rm \Delta p}({\bf r}')=4\sum_{{\bf R '},o}{\rm Re}
\left[\langle \bar \Psi_{o}^{(0)}|\Delta{\bf J}^{\rm p}_{\bf R'}({\bf
r}'){\cal G}(\varepsilon_o) \bar H^{(1)}| \bar
\Psi_{o}^{(0)}\rangle-\langle \bar \Psi_{o}^{(0)}|\Delta{\bf J}^{\rm
p}_{\bf R'}({\bf r}'){\cal G}(\varepsilon_o) \frac{{\bf B}\times {\bf
R'}}{2c}\cdot {\bf v}| \bar \Psi_{o}^{(0)}\rangle\right],
\label{jpbis}
\end{equation}
where ${\bf v}= 1/i[{\bf r},\bar H^{(0)}]$ is the velocity operator.
Now the translational invariance of ${\bf j}^{(1)}_{\rm bare}({\bf
r}')$ and ${\bf j}^{(1)}_{\rm \Delta p}({\bf r}')$ is more explicit,
since on translation both ${\bf B}\times {\bf r}'\cdot {\bf v}/2c$
and ${\bf B}\times {\bf R'}\cdot {\bf v}/2c$ generate an extra term
equal to ${\bf B}\times {\bf t}\cdot {\bf v}/2c$, as does $\bar
H^{(1)}$, if, after the translation, we rewrite $\bar H^{(1)}$ in terms
of the variable of the translated coordinate system $({\bf r}-{\bf
t})$.

\section{Current response in extended systems}
\label{current-extend}

In Section \ref{pspinb} we developed a theory for a system containing
a single augmentation region located at the origin, and then later for
several augmentation regions. We must now check that our results are
still useful in situations involving an infinite number of these
augmentation regions, as is the case in the solid state.
 
The expression for ${\bf j}^{(1)}_{\rm \Delta d}({\bf r}')$ given by
Eq. (\ref{jd}) can be straightforwardly applied to solid state
calculations. But, the contributions to the all-electron current ${\bf
j}^{(1)}_{\rm bare}({\bf r}')$ and ${\bf j}^{(1)}_{\rm \Delta p}({\bf
r}')$ given in Eq. (\ref{jbarebis}) and Eq. (\ref{jpbis}) involve
expectation values of the position operator. As these are not
generally defined in an extended system, one might worry that
Eqs. (\ref{jbarebis}) and (\ref{jpbis}) are not valid. However, if
they are rewritten in the following way:
\begin{equation}
\label{jbare-rewrite}
{\bf j}^{(1)}_{\rm bare}({\bf r}') = \frac{2}{c}\sum_o {\rm Re} \left[
\bracket{\bar \Psi_{o}^{(0)}}{{\bf J}^{\rm p}({\bf r}'){\cal
G}(\varepsilon_o)\left(({\bf r}-{\bf r}')\times{\bf p}+\sum_{\bf
R}({\bf R}-{\bf r}')\times{\bf v}^{\rm nl}_{\bf R}\right)\cdot{\bf
B}}{\bar \Psi_{o}^{(0)}}\right]
\end{equation}
and
\begin{equation}
\label{jdp-rewrite}
{\bf j}^{(1)}_{\rm \Delta p}({\bf r}') = \frac{2}{c}\sum_{{\bf R}',o}
{\rm Re} \left[ \bracket{\bar \Psi_{o}^{(0)}}{\Delta{\bf J}^{\rm
p}_{{\bf R}'}({\bf r}'){\cal G}(\varepsilon_o)\left(({\bf r}-{\bf
R}')\times{\bf p}+\sum_{\bf R}({\bf R}-{\bf R}')\times{\bf v}^{\rm
nl}_{\bf R}\right)\cdot{\bf B}}{\bar \Psi_{o}^{(0)}}\right]
\end{equation}
then it becomes clear that they are indeed well defined. The Green
Function operator ${\cal G}(\varepsilon_o)$, in an insulator, and both
the paramagnetic augmentation operator $\Delta{\bf J}^{\rm p}_{{\bf
R}'}({\bf r}')$ and the non-local pseudopotential operator ${\bf
v}^{\rm nl}_{\bf R}$ are short ranged.\cite{footnote}
This ensures that contributions to the
current response for large values of $({\bf r}-{\bf r}')$, $({\bf
R}-{\bf r}')$, $({\bf r}-{\bf R}')$, or $({\bf R}-{\bf R}')$ in
Eqs. (\ref{jbare-rewrite}) and (\ref{jdp-rewrite}) vanish.

\section{Current response in infinitely periodic systems}
\label{current-pbc-extend}

The expressions given above are valid for any extended system.
However, the only such computationally tractable systems are those
exhibiting translational symmetry, or infinitely periodic systems. We
now develop the equations making this translational symmetry explicit,
by writing the electronic states as Bloch functions,
$\ket{\bar\Psi_{n,{\bf k}}^{(0)}}=e^{i{\bf k}\cdot{\bf r}}\ket{\bar
u^{(0)}_{n,{\bf k}}}$, where ${\bf k}$ is a reciprocal space vector
within the first Brillouin zone and the corresponding eigenvalues are
$\varepsilon_{n,{\bf k}}$. The cell-periodic function $\braket{{\bf
r}}{\bar u^{(0)}_{n,{\bf k}}}$ is normalized within the unit cell.

In order to take full advantage of this translational symmetry we
first define the functions ${\bf S}_{\rm bare} ({\bf r}',q)$ and ${\bf
S}_{\rm \Delta p} ({\bf r}',q)$ as:
\begin{equation}
\label{Sbare}
{\bf S}_{\rm bare} ({\bf r}',q) =
\frac{2}{c}\sum_{i=x,y,z}\sum_{o}{\rm Re} \left[
\frac{1}{i}\bracket{\bar \Psi_{o}^{(0)}}{{\bf J}^{\rm p}({\bf
r}'){\cal G}(\varepsilon_o){\bf B}\times\hat{\bf
u}_i\cdot\left(e^{iq\hat{\bf u}_i\cdot({\bf r}-{\bf r}')}{\bf
p}+\sum_{\bf R}e^{iq\hat{\bf u}_i\cdot({\bf R}-{\bf r}')}{\bf v}^{\rm
nl}_{\bf R}\right)}{\bar \Psi_{o}^{(0)}}\right]
\end{equation}
\begin{equation}
\label{Sdp}
{\bf S}_{\rm \Delta p} ({\bf r}',q) =
\frac{2}{c}\sum_{i=x,y,z}\sum_{{\bf R}',o}{\rm Re} \left[
\frac{1}{i}\bracket{\bar \Psi_{o}^{(0)}}{\Delta{\bf J}^{\rm p}_{{\bf
R}'}({\bf r}'){\cal G}(\varepsilon_o){\bf B}\times\hat{\bf
u}_i\cdot\left(e^{iq\hat{\bf u}_i\cdot({\bf r}-{\bf R}')}{\bf
p}+\sum_{\bf R}e^{iq\hat{\bf u}_i\cdot({\bf R}-{\bf R}')}{\bf v}^{\rm
nl}_{\bf R}\right)}{\bar \Psi_{o}^{(0)}}\right],
\end{equation}
where the $\hat{\bf u}_i$ are unit vectors in the three Cartesian
directions. We can then write
\begin{equation}
\label{jbare-diff}
{\bf j}^{(1)}_{\rm bare}({\bf r}') =
\lim_{q\rightarrow0}\frac{1}{2q}\left[{\bf S}_{\rm bare} ({\bf
r}',q)-{\bf S}_{\rm bare} ({\bf r}',-q)\right]
\end{equation}
\begin{equation}
\label{jdp-diff}
{\bf j}^{(1)}_{\rm \Delta p}({\bf r}') =
\lim_{q\rightarrow0}\frac{1}{2q}\left[{\bf S}_{\rm \Delta p} ({\bf
r}',q)-{\bf S}_{\rm \Delta p} ({\bf r}',-q)\right].
\end{equation}
This can be seen to be correct by expanding the exponentials in
Eqs. (\ref{Sbare}) and (\ref{Sdp}) as $e^{iq\hat{\bf u}_i\cdot{\bf
x}}=1+iq\hat{\bf u}_i\cdot{\bf x}+O((qx)^2)$, taking the limits in
Eqs. (\ref{jbare-diff}) and (\ref{jdp-diff}) and comparing to
Eqs. (\ref{jbare-rewrite}) and (\ref{jdp-rewrite}). The limits taken using
the expanded exponentials are valid since only finite values of ${\bf
x}$ contribute to the total current (as established in Section
\ref{current-extend}).

The description of the electronic states as Bloch functions allows us
to approximate the summations over the infinite number of occupied
states in Section \ref{current-extend} as finite summations over ${\bf
k}$-dependent quantities. The ${\bf k}$-dependent Green function is,
\begin{equation}
{\cal G}_{\bf k}(\varepsilon)=\sum_{e}\frac{|\bar u_{e,{\bf
k}}^{(0)}\rangle \langle \bar u_{e,{\bf k}}^{(0)}|}{
\varepsilon-\varepsilon_{e,{\bf k}}}.
\end{equation}
A consequence of re-expressing the current contributions in terms of
${\bf S}_{\rm bare} ({\bf r}',q)$ and ${\bf S}_{\rm \Delta p} ({\bf
r}',q)$ is that we must evaluate several quantities at ${\bf k}$ and
${\bf k}+{\bf q}$ simultaneously. For example, the usual form of the
${\bf k}$-dependent non-local pseudopotential operator is generalized:
\begin{equation}
V^{\rm nl}_{{\bf k},{\bf k}'} = \sum_{\bf \tau}\sum_{n,m}\ket{\tilde
p^{\bf k}_{{\bf \tau},n}}a^{\bf \tau}_{n,m}\bra{\tilde p^{{\bf k}'}_{{\bf
\tau},m}}.
\end{equation}
This operator acts on Bloch functions at ${\bf k}$ to the left, and
${\bf k}'$ to the right. For $\bf k=k'$, $V^{\rm nl}_{{\bf k},{\bf k}'}$ 
coincides with the $\bf k$-dependent non-local pseudopotential operator, 
implemented in the plane-wave pseudopotential codes.
The ${\bf k}$-dependent projectors in terms
of $\ket{\tilde p_{{\bf R},n}}$, the real space projectors, are given
by
\begin{equation}
\ket{\tilde p^{\bf k}_{{\bf \tau},n}} = \sum_{\bf L}e^{-i{\bf
k}\cdot({{\bf r}-{\bf L}-{\bf \tau}})}\ket{\tilde p_{{\bf L}+{\bf
\tau},n}},
\end{equation}
where the ${\bf L}$ are lattice vectors and the ${\bf \tau}$ are the
internal co-ordinates of the atoms. We arrive at analogous expressions
for both the velocity operator,
\begin{equation}
{\bf v}_{{\bf k},{\bf k}'} = -i{\bf \nabla}+{\bf k}' +
\frac{1}{i}[{\bf r},V^{\rm nl}_{{\bf k},{\bf k}'}],
\end{equation}
and the paramagnetic current operator,
\begin{equation}
{\bf J}^{\rm p}_{{\bf k},{\bf k}'}({\bf r}')=-\frac{(-i{\bf
\nabla}+{\bf k})\ket{{\bf r}'}\bra{{\bf r}'}+\ket{{\bf r}'}\bra{{\bf
r}'}(-i{\bf \nabla}+{\bf k}')}{2}.
\end{equation}
Combining the above we arrive at a compact expression for ${\bf S}_{\rm
bare}({\bf r}',q)$:
\begin{equation}
\label{jbare-uk}
{\bf S}_{\rm bare}({\bf r}',q) = \frac{2}{cN_{\rm
k}}\sum_{i=x,y,z}\sum_{o,{\bf k}}{\rm Re}\left[
\frac{1}{i}\bracket{\bar u^{(0)}_{o,{\bf k}}}{{\bf J}^{\rm p}_{{\bf
k},{\bf k}+{\bf q}_i}({\bf r}'){\cal G}_{{\bf k}+{\bf
q}_i}(\varepsilon_{o,{\bf k}}){\bf B}\times\hat{\bf u}_i\cdot{\bf
v}_{{\bf k}+{\bf q}_i,{\bf k}}}{\bar u^{(0)}_{o,{\bf k}}}\right],
\end{equation}
where ${\bf q}_i=q\hat{\bf u}_i$ and $N_{\rm k}$ is the number of ${\bf
k}$-points included in the summation. Similarly, by also defining:
\begin{equation}
\Delta{\bf J}^{\rm p}_{{\bf L},{\bf \tau},{\bf k},{\bf k}'}({\bf
r}')=\sum_{n,m}|\tilde p^{\bf k}_{{\bf \tau},n}\rangle
\left[\langle \phi_{{\bf L}+{\bf \tau},n}| {\bf J}^{\rm p}({\bf
r}')|\phi_{{\bf L}+{\bf \tau},m}\rangle - \langle \tilde \phi_{{\bf
L}+{\bf \tau},n}| {\bf J}^{\rm p}({\bf r}') |\tilde\phi_{{\bf L}+{\bf
\tau},m}\rangle\right] \langle\tilde p^{{\bf k}'}_{{\bf \tau},m}|,
\end{equation}
the expression for the paramagnetic augmentation term is:
\begin{equation}
\label{jdp-uk}
{\bf S}_{\rm \Delta p}({\bf r}',q) = \frac{2}{cN_{\rm
k}}\sum_{i=x,y,z}\sum_{{\bf L},{\bf \tau},o,{\bf k}}{\rm Re}\left[
\frac{1}{i}\bracket{\bar u^{(0)}_{o,{\bf k}}}{\Delta {\bf J}^{\rm
p}_{{\bf L},{\bf \tau},{\bf k},{\bf k}+{\bf q}_i}({\bf r}'){\cal
G}_{{\bf k}+{\bf q}_i}(\varepsilon_{o,{\bf k}}){\bf B}\times\hat{\bf
u}_i\cdot{\bf v}_{{\bf k}+{\bf q}_i,{\bf k}}}{\bar u^{(0)}_{o,{\bf
k}}}\right].
\end{equation}
These expressions for ${\bf S}_{\rm bare}({\bf r}',q)$ and ${\bf
S}_{\rm \Delta p}({\bf r}',q)$ allow the evaluation of the
all-electron current response through the Eqs. (\ref{jbare-diff}),
(\ref{jdp-diff}) and (\ref{jd}).

\section{Summary of Approaches}
\label{summary}

There are three different approaches that we could take in the
calculation of the first order current response to a uniform external
applied magnetic field. If the current response in an extended
periodic system is required, then the approach described in Section
\ref{current-pbc-extend} must be taken. In this case, the expressions
given in Eqs. (\ref{jbare-diff}), (\ref{jdp-diff}), (\ref{jbare-uk}),
(\ref{jdp-uk}) and (\ref{jd}) are evaluated, and it is referred to as
the ``crystal approach''. The total current response in a finite
system can be calculated using Eqs. (\ref{jbare}), (\ref{jp}), and
(\ref{jd}). This approach is referred to as the ``molecular
approach''. Alternatively, using the results of the generalized
$f$-sum rule, Eqs. (\ref{jbarebis}), (\ref{jpbis}), and (\ref{jd}) can
be used. This is the ``molecular sum rule approach''.  Setting ${\cal
T}_{\bf B}={\bf 1}$ in the GIPAW formalism, i.e. in the all-electron
case, the ``crystal approach'' becomes equivalent to the MPL
method,\cite{mauri96II} the ``molecular approach'' becomes equivalent
to the single gauge method (Eq. (3) of Ref. \onlinecite{gregor99}),
and the ``molecular sum rule approach'' becomes equivalent to the
continuous set of gauge transformation method\cite{CSGT} (CSGT) with
the $\bf d(r)=r$ gauge function (Eq. (8) of
Ref. \onlinecite{gregor99}).

The ``crystal approach'' can be used to calculate molecular properties
through the use of large supercells.  If the generalized $f$-sum rule
holds, then the results obtained by each of the three approaches
should be equivalent. This is demonstrated in Section \ref{testGIPAW}.
If the generalized $f$-sum rule does not hold well (for example, if
the basis set used is far from completeness as is the case for the
atomic-orbital basis sets used in most quantum chemical
calculations\cite{gregor99}), then the ``crystal approach'' and the
``molecular sum rule approach'' will still give the same
results. However, the results obtained using the ``molecular
approach'' will be different.  In particular, we expect that the
``molecular approach'' will require a much larger atomic-orbital basis
set to converge the NMR chemical shifts than the other two methods, as
it has been proved to be the case for all-electron
Hamiltonians.\cite{CSGT} This is because the two terms in
Eq. (\ref{jbare}) (as well as the two terms in Eq. (\ref{jp})) of the
``molecular approach'' converge at different rates with respect to the
completeness of the basis set.\cite{CSGT,gregor99}

\section{Calculation of NMR Chemical Shifts}
\label{calc-nmrshifts}

It is important to show that the GIPAW method is a practical approach
to the calculation of NMR chemical shifts. We have therefore
implemented the method into a parallelized plane-wave pseudopotential
electronic structure code.\cite{mcp:rev} Such codes
self consistently calculate the ground state electronic
structure. Specifically, the self consistent Hamiltonian $\bar
H^{(0)}$ and the corresponding wavefunctions $| \bar
\Psi_{n}^{(0)}\rangle$ that appear in the above expressions are
obtained. In this section we outline the features of the
implementation that are specific to the GIPAW method, and not to the
pseudopotential method in general. The plane-wave pseudopotential
method is most naturally suited to the ``crystal approach'' for the
calculation of NMR chemical shifts. However, we also implemented both
molecular methods in our plane-wave code, for completeness. This is
described in Section \ref{finite-pbc}.

\subsection{Application of the Green function}

There are several points at which first order wavefunctions of the
form,
\begin{equation}
|\bar \Psi_{n}^{(1)}\rangle = {\cal G}(\varepsilon_n)\bar H^{(1)}|
 \bar \Psi_{n}^{(0)}\rangle
\label{greenapp}
\end{equation}
must be evaluated. The Green function ${\cal G}(\varepsilon)$ is
given by,
\begin{equation}
{\cal G}(\varepsilon)=\sum_{e}\frac{|\bar\Psi_{e}^{(0)}\rangle \langle
\bar \Psi_{e}^{(0)}|}{ \varepsilon-\varepsilon_e },
\label{greenfn2}
\end{equation}
and a naive approach would require the explicit summation over all empty
states. This is unnecessarily arduous. We can multiply
Eq. (\ref{greenapp}) through by $(\varepsilon_n-\bar H^{(0)})$. If we
then write ${\cal Q} = \sum_e|\bar\Psi_{e}^{(0)}\rangle \langle\bar
\Psi_{e}^{(0)}| = 1 - \sum_o|\bar\Psi_{o}^{(0)}\rangle \langle\bar
\Psi_{o}^{(0)}|$, where the sums over $o$ and $e$ are over the
occupied and empty states respectively, we obtain:
\begin{equation}
(\varepsilon_n-\bar H^{(0)})|\bar \Psi_{n}^{(1)}\rangle = {\cal Q}\bar
 H^{(1)}| \bar \Psi_{n}^{(0)}\rangle
\end{equation}
This is a linear system involving only the occupied states, and can be
solved using a conjugate gradient minimization scheme,\cite{conjg} as 
in Ref. \onlinecite{mauri96II}.
This approach ensures that our method is comparable in computational
cost to the calculation of the ground-state electronic structure.

\subsection{The velocity operator}

The velocity operator ${\bf v}=\frac{1}{i}[{\bf r},\bar H^{(0)}]$
appears in various guises throughout the relevant expressions
above. The velocity operator may also be written as the first
derivative of the ${\bf k}$-dependent Hamiltonian with respect to
${\bf k}$. The term related to the kinetic energy is straightforward
to evaluate, and is simply the momentum operator. The term due to the
non-local potential, which is defined numerically, is best obtained
numerically. In our implementation we simply take the appropriate
numerical derivative of the ${\bf k}$-dependent non-local potential
operators. The derivatives are evaluated by calculating the non-local
potential at, say, ${\bf k}$ and ${\bf k+q}$, where ${\bf q}$ is
chosen to be small enough that the resulting numerical derivative is
accurate, but not so small so as to introduce numerical noise.

\subsection{The crystal approach}

The ``crystal approach'' requires that the limits of
Eqs. (\ref{jbare-diff}) and \ref{jdp-diff} are evaluated. These are,
in effect, similar to the numerical derivatives which must be take in
reciprocal space in order to evaluate the velocity operators. And in
practice we take the same value for the reciprocal space step size in
both cases.  The same considerations apply. The step should be chosen
to be small enough that the resulting limit is accurately
approximated, but not so small that numerical noise dominates. A
typical value is 0.01~Bohr$^{-1}$.

\subsection{Finite systems in periodic boundary conditions}
\label{finite-pbc}

Both the ``molecular approach'' and the ``molecular sum rule
approach'' were implemented. The major difference between these
approaches and the ``crystal approach'', from a computational
perspective, is that the reciprocal space numerical derivative is
replaced by a direct application of the position operator to the
wavefunctions. Clearly, the position operator is not defined within
periodic boundary conditions. But we can treat it approximately by
constructing a periodic saw-tooth like function (in practice we build
the function in reciprocal space). Near the center of the simulation
cell, or about wherever the saw-tooth is centered, this operator
approximates the position operator. This approximation improves as the
size of the simulation cell is increased, and for good results the
magnitude of the induced current should be small on the surface where
the saw-tooth function changes sign.

\subsection{From the current to the NMR chemical shifts}

The GIPAW approach separates the contributions to the current response
into a bare term, ${\bf j}^{(1)}_{\rm bare}({\bf r})$, and two
correction terms, the paramagnetic and diamagnetic corrections, ${\bf
j}^{(1)}_{\rm \Delta p}({\bf r})$ and ${\bf j}^{(1)}_{\rm \Delta
d}({\bf r})$ respectively.  To compute the NMR chemical shifts, using
Eq. (\ref{bs2}), the induced magnetic field,
${\bf B}^{(1)}_{\rm in}({\bf R})$, must be evaluated at each nuclear
position $\bf R$.  In principle, one could combine the three current
contributions and obtain ${\bf B}^{(1)}_{\rm in}({\bf R})$ from the
total current using Eq. (\ref{bs}).  We use a different approach. We
take advantage of the linearity of Eq. (\ref{bs}), and we solve it for
each of the three current contributions, obtaining a bare induced
field ${\bf B}^{(1)}_{\rm bare}({\bf R})$, a paramagnetic correction
field, ${\bf B}^{(1)}_{\rm \Delta p}({\bf R})$, and a diamagnetic
correction field, ${\bf B}^{(1)}_{\rm \Delta d}({\bf R})$.

To compute the correction fields, we suppose that just the correction
currents, ${\bf j}^{(1)}_{\rm \Delta p}({\bf r})$ and ${\bf
j}^{(1)}_{\rm \Delta d}({\bf r})$, within the augmentation region
$\Omega_{\bf R}$ contribute to ${\bf B}^{(1)}_{\rm \Delta p}({\bf R})$
and ${\bf B}^{(1)}_{\rm \Delta d}({\bf R})$ at the nuclear position
$\bf R$.  Using this on-site approximation, combining Eqs. (\ref{bs})
and (\ref{jd}), we obtain:
\begin{equation}
{\bf B}^{(1)}_{\rm \Delta d}({\bf R})=
2\sum_{o,n,n'}\langle \bar \Psi_{o}^{(0)}|
\tilde p_{{\bf R},n}\rangle {\bf e}^{\bf R}_{m,n}
\langle
\tilde p_{{\bf R},m}| \bar
\Psi_{o}^{(0)}\rangle,
\end{equation}
where
\begin{equation}
{\bf e}^{\bf R}_{m,n}=
\langle \phi_{{\bf R},n}|
\frac{({\bf R}-{\bf r}) \times [{\bf B}\times({\bf
R}-{\bf r})]}{2c^2|{\bf R}-{\bf r}|^3}
|\phi_{{\bf R},m}\rangle
-
\langle \tilde \phi_{{\bf R},n}|
\frac{({\bf R}-{\bf r}) \times [{\bf B}\times({\bf
R}-{\bf r})]}{2c^2|{\bf R}-{\bf r}|^3}
|\tilde\phi_{{\bf R},m}\rangle,
\end{equation}
The coefficients ${\bf e}^{\bf R}_{m,n}$ depend only on the atomic
species, and need only be calculated once.  Similarly, within the
on-site approximation, by combining Eq. (\ref{bs}) with the equations for
the ${\bf j}^{(1)}_{\rm \Delta p}({\bf r})$ correction current, we
obtain expressions for the paramagnetic correction field, ${\bf
B}^{(1)}_{\rm \Delta p}({\bf R})$, which depend linearly on the
coefficients ${\bf f}^{\bf R}_{m,n}$,
\begin{equation}
\label{f-para}
{\bf f}^{\bf R}_{m,n}= \langle \phi_{{\bf R},n}| \frac{{\bf
L_R}}{|{\bf r}-{\bf R}|^3} |\phi_{{\bf R},m}\rangle - \langle \tilde
\phi_{{\bf R},n}| \frac{{\bf L_R}}{|{\bf r}-{\bf R}|^3}
|\tilde\phi_{{\bf R},m}\rangle,
\end{equation}
where ${\bf L_R}=({\bf r}-{\bf R})\times{\bf p}$ is the angular
momentum operator evaluated with respect to the atomic site $\bf R$.
Again, the coefficients ${\bf f}^{\bf R}_{m,n}$ depend only on the
atomic species, and need only be evaluated once.
 
To compute the bare induced field, ${\bf B}^{(1)}_{\rm bare}({\bf
R})$, we Fourier transform Eq. (\ref{bs}) and ${\bf j}^{(1)}_{\rm
bare}({\bf r}')$ into reciprocal space.  The induced magnetic field
can then be simply evaluated as,
\begin{equation}
\label{bsG}
{\bf B}^{(1)}_{\rm bare}({\bf G})= \frac{4 \pi}{c} \frac{i{\bf G}
\times {\bf j}^{(1)}_{\rm bare}({\bf G})}{G^2},
\end{equation}
where $\bf G$ is a reciprocal lattice vector. We subsequently obtain
${\bf B}^{(1)}_{\rm bare}({\bf R})$ by a slow (since we only need the
results at a few points in space) Fourier transform at the nuclear
positions $\bf R$.

For $\bf G=0$, Eq. (\ref{bsG}) can not be applied.  Indeed the $\bf
G=0$ component of the induced magnetic field is not a bulk
property.\cite{mauri96II} The $\bf G=0$ component of the induced field
is affected by the surface currents which appear on the surface of the
sample.  In particular, its value depends on the the shape of the
sample, and is determined by macroscopic magnetostatics.  Following
the experimental convention, we assume a spherical sample in our
calculations, for which:
\begin{equation}
{\bf B}^{(1)}_{\rm in}({\bf G=0})=\frac{8\pi}{3}  {\tensor \chi}{\bf B},
\end{equation}
where ${\tensor \chi}$ is the macroscopic magnetic
susceptibility.\cite{mauri96II} To be consistent with the on-site
approximation for the correction currents, we should not take into
account the contribution of ${\bf j}^{(1)}_{\rm \Delta p}({\bf r})$
and ${\bf j}^{(1)}_{\rm \Delta d}({\bf r})$ to ${\bf B}^{(1)}_{\rm
in}({\bf G=0})$, and so use:
\begin{equation}
{\bf B}^{(1)}_{\rm in}({\bf G=0})=\frac{8\pi}{3}  
{\tensor \chi_{\rm bare}}{\bf B} ,
\end{equation}
where ${\tensor \chi_{\rm bare}}$ is the contribution to the 
macroscopic susceptibility coming from the bare current 
${\bf j}^{(1)}_{\rm bare}({\bf r})$.
Within the ``crystal approach'', 
we use the following ansatz for ${\tensor \chi_{\rm bare}}$:
\begin{equation}
\label{chi}
{\tensor \chi_{\rm bare}}=\lim_{q\rightarrow0}
\frac{{\tensor F}(q)-2{\tensor F}(0)+{\tensor F}(-q)}{q^2},
\end{equation}
where $F_{ij}(q)=(2-\delta_{ij})Q_{ij}(q)$, $i$ and $j$ are Cartesian
indices,
\begin{equation}
{\tensor Q}(q) = -\frac{1}{c^2 N_{\rm k}
V_c}\sum_{i=x,y,z}\sum_{o,{\bf k}}{\rm Re}\left[ \bracket{\bar
u^{(0)}_{o,{\bf k}}} { \hat{\bf u}_i\times(-i{\bf \nabla}+{\bf k} )
{\cal G}_{{\bf k}+{\bf q}_i}(\varepsilon_{o,{\bf k}})\hat{\bf
u}_i\times{\bf v}_{{\bf k}+{\bf q}_i,{\bf k}}}{\bar u^{(0)}_{o,{\bf
k}}}\right],
\end{equation}
and $V_{\rm c}$ is the unit cell volume.  In support of this ansatz,
one can show that, when ${\cal T}_{\bf B}={\bf 1}$, i.e.  in the
all-electron case, the definition of $\tensor \chi_{\rm bare}$,
Eq. (\ref{chi}) becomes equal to the expression for the calculation of
the all-electron macroscopic magnetic susceptibility, as derived in
Ref. \onlinecite{mauri96I}.  

\subsection{Projectors}

In our implementation, we use norm-conserving Troullier-Martins
pseudopotentials\cite{tm:vps} with single projectors for each angular
momentum channel. As a result, the argument in Section \ref{oneregion}
holds and the $b^{(1)}_{n,m}$ terms are zero. However, in contrast to
what Van de Walle and Bl\"ochl found for the calculation of hyperfine
parameters,\cite{vandewalle93} we found that a minimum of two
projectors per channel were required to ensure good transferability of
the GIPAW current corrections. Otherwise, the projectors are
constructed as described in Ref. \onlinecite{vandewalle93}, except
that we choose a polynomial step function $f(r)$ so that the
pseudowavefunctions are cut off smoothly at some distance less than
the pseudopotential core radius.

\section{Numerical tests of the GIPAW method}
\label{testGIPAW}

\subsection{Comparison with IGAIM results}

Quantum chemical approaches have long been able to predict the NMR
chemical shifts of small molecules, and one of the most widely used is
the GAUSSIAN94\cite{gaussian94} quantum chemical code. Gregor \emph{et
al}\cite{gregor99} used this code to optimize the geometry and
calculate the isotropic chemical shift of a selection of small
molecules using both the GIAO and IGAIM methods.\cite{igaimgaussian94}
We compare our GIPAW results (all chemical shifts reported here have
been calculated within the Local Density Approximation\cite{ca:lda})
to the IGAIM results for several of these molecules (using exactly the
same relaxed geometries) in Table \ref{IGAIMvsGIPAW}. 
The total isotropic chemical shifts computed with  GIPAW agree very well
in all cases with the  GAUSSIAN94 results. 

The GIPAW results presented in Table \ref{IGAIMvsGIPAW} were evaluated
using the ``crystal approach'', but results obtained using the
molecular approaches differ typically by less than 0.1
parts-per-million (ppm) in sufficiently large simulation cells, as
demonstrated in Table \ref{IGAIMvs3xGIPAW}. 

The GIPAW results are converged to the 0.1 ppm level using a
plane-wave cut-off of 100 Rydbergs, a super-cell volume of 6000
Bohr$^3$ and a $2\times2\times2$ Monkhorst-Pack {\bf k}-point
grid.\cite{monkhorst} The states indicated in Table \ref{IGAIMvsGIPAW}
were treated as core states in the pseudopotential calculations.  The
core contribution to the GIPAW chemical shifts is assumed to be
constant (following the observations of Gregor \emph{et
al}\cite{gregor99}), and evaluated in an all-electron atomic code.
For hydrogen a pseudisation core radius of 1.2 Bohr was used and only
the s-channel was augmented.  As a result, since the paramagnetic
correction term is proportional to the angular momentum of the
augmentation channel (see Eq. \ref{f-para}), only the bare and
diamagnetic correction terms contribute to the total isotropic
chemical shifts. There is no core contribution for hydrogen. For the
carbon shifts the s- and p-channels were augmented and a core radius
of 1.6 Bohr used in the generation of the pseudopotential. For silicon
and phosphorus the d-channel was also augmented and core radii of 2.0
Bohr used in both cases. Gregor \emph{et al} attempted to converge the
chemical shifts with respect to their localized basis set size, and
the convergence appears to be to the 1 ppm level for the carbon and
silicon shifts (see Fig. 2 of Ref. \onlinecite{gregor99}). However,
the convergence appears to be less complete for the phosphorus
shifts. It is just these chemical shifts for which the GIPAW and IGAIM
results differ the most (although the errors as a fraction of the
range of the chemical shifts are similar for all nuclei).  While the
diamagnetic correction term is found to be rigid with respect to
chemical environment, both the bare and paramagnetic correction terms
are found to be strongly dependent on the system. The correction terms
introduced by the GIPAW approach are therefore seen to be important
even in the prediction of relative chemical shifts, and the rigid
nature of the core contribution is reconfirmed.

In Table \ref{psp-methane} we examine the robustness of the GIPAW
method with respect to pseudopotentials used. A variety of
Troullier-Martins pseudopotentials\cite{tm:vps}, with core radii
ranging from 1.2 to 1.8 Bohr, were used to calculate the NMR chemical
shift for carbon in methane. While the bare contribution to the
chemical shift is observed to change by over 10 ppm, the total shifts,
including the GIPAW correction terms, are constant to within 1
ppm. There is virtually no difference in the total shifts between
potentials with core radii of 1.2 and 1.4 Bohr.

\subsection{Comparison with all-electron plane-wave results for diamond}

As the GIPAW method presented here is, to the authors' knowledge, the
only approach available for the calculation of all-electron NMR
chemical shifts in solids, a truly independent validation is not
possible. However, by constructing a suitable pseudopotential and
taking a high enough plane-wave cut-off energy we are able to compare
with essentially all-electron results --- in which all the electrons
in the chosen system are considered to be valence electrons. In this
way we can check the corrections to the conventional pseudopotential
results. Obviously, such calculations are computationally intensive
due to the extremely large number of plane-waves required to reach
convergence. We therefore choose diamond as our example periodic
system. Carbon is sufficiently light that an all-electron
plane-wave calculation is possible, and the diamond structure has a
very small primitive unit cell and a high degree of symmetry. The
$1s$, $2s$, and $2p$ electrons are all considered to be valence
electrons and we construct a purely local Troullier-Martin
\cite{tm:vps} pseudopotential with a core radius of 0.4 Bohr radii.

Table \ref{sigma-diamond} compares the results of a GIPAW
pseudopotential calculation (the $1s$ electrons are treated as core
electrons, and a core radius of 1.6 Bohr radii used) and the
all-electron plane-wave calculation obtained with the purely local
Troullier-Martin pseudopotential. The contributions can be separated
into core and valence terms in a gauge invariant way, as shown in
Ref. \onlinecite{gregor99}.  Thus, in the case of the all-electron
result we performed two calculations of the chemical shift after
achieving self-consistency, once taking into account all the
electrons, and a second time excluding the valence electrons from the
calculation of the chemical shift. The valence term presented is the
difference between these two results. We present the all-electron
results at two plane-wave cut-offs --- 800 and 1400 Rydbergs and a
$10\times10\times10$ Monkhorst-Pack ${\bf k}$-point grid. All the
contributions to the chemical shifts are converged to within a
part-per-million. The valence contributions of the GIPAW and
all-electron results differ by only 1.39 ppm which may be attributed
to the slight uncorrected pseudisation error that remains in the
all-electron result. We have confidence that if the core radius were
reduced to less than 0.4 Bohr radii the difference between the results
of the two approaches would decrease. The GIPAW pseudopotential result
is expected to be closer to the true all-electron NMR chemical shift.

\section{Conclusions}

We have presented an \emph{ab initio} theory for the evaluation of NMR
chemical shifts in both finite and infinitely periodic systems. We
have correctly treated the complications introduced due to the use of
pseudopotentials, and so, in contrast to the original implementation
of the MPL approach,\cite{mauri96II}
we are not restricted to the calculation of the chemical shifts for
light elements. We introduced an extension to the Projector
Augmented-Wave method which is valid for systems in non-zero uniform
magnetic fields, the Gauge Including Projector Augmented-Wave
method. 

Our implementation of GIPAW into a parallelized plane-wave
pseudopotential code allows the calculation of NMR chemical shifts in
large, low symmetry extended systems. We expect that the methodology
will prove useful in the calculation of other magnetic properties.  
Our work also suggests that the
implementation of GIPAW into quantum chemical approaches would lead to
a considerable improvement in their efficiency for the calculation of
NMR chemical shifts for heavy elements. 

\section*{Acknowledgments}

Some of the calculations were performed at the IDRIS supercomputer
center of the CNRS.  CJP would like to thank the Deutsche
Forschungsgemeinschaft for funding under grant PI 398/1-1, and the
Universit\'{e} Paris 6 and the Universit\'{e} Paris 7 for the support
during his stay in Paris.

\appendix

\section{The Generalized $f$-sum rule}
\label{Asum-rule}

The generalized $f$-sum rule holds for any pair of hermitian operators
${\cal O}$ and ${\cal E}$, where ${\cal O}$ and ${\cal E}$ are
respectively odd and even on time reversal, i.e.:
\begin{equation} 
\langle \phi|{\cal O}|\phi'\rangle=-\langle \phi'| {\cal O}|\phi\rangle
\end{equation}
and 
\begin{equation}
\langle\phi| {\cal E}|\phi'\rangle=\langle \phi'| {\cal E}|\phi\rangle
\end{equation}
for any $|\phi\rangle$ and $|\phi'\rangle$ such that $\braket{\bf
r}{\phi}$ and $\braket{\bf r}{\phi'}$ are real. It is straightforward
to verify that $\bf p$, $\bf L$, $\bf v$, ${\bf v}_{\bf R}^{\rm nl}$,
${\bf J}^p({\bf r}')$, and $\Delta{\bf J}^p_{\bf {\bf R}}({\bf r}')$
are odd, and that ${\bf r}$ and operators that are a function of
${\bf r}$ are even.  To derive the sum rule, we consider the quantity
\begin{equation}
{ s}=-4\sum_{o}{\rm Re}\left[\langle \bar \Psi_{o}^{(0)}|{\cal O}{\cal
G}(\varepsilon_0)\frac{1}{i}[{\cal E},\bar H^{(0)}]| \bar
\Psi_{o}^{(0)}\rangle\right].
\end{equation}
The sums over $o$ and $o'$ (below) run over the occupied orbitals, and
those over $e'$ over the empty ones.  Using the fact that $\bar
H^{(0)}| \bar \Psi_{k}^{(0)}\rangle= \varepsilon_k| \bar
\Psi_{k}^{(0)}\rangle$, Eq. (\ref{greenfn}) and $\sum_{e'}|\hat
\Psi_{e'}\rangle\langle\bar\Psi_{e'}|={\bf
1}-\sum_{o'}|\bar\Psi_{o'}\rangle\langle\bar\Psi_{o'}|$ the expression
for $s$ may be rewritten as,
\begin{equation}
{ s}=-4\sum_o{\rm Re}\left[\frac{1}{i}\langle \bar \Psi_{o}^{(0)}|
{\cal O} {\cal E}| \bar \Psi_{o}^{(0)}\rangle\right]+4\sum_{o,o'}{\rm
Re}\left[\frac{1}{i}\langle \bar \Psi_{o}^{(0)}| {\cal O}| \bar
\Psi_{o'}^{(0)}\rangle\langle \bar \Psi_{o'}^{(0)}|{\cal E}| \bar
\Psi_{o}^{(0)}\rangle\right].
\label{spartial}
\end{equation}
Since the eigenstates $ | \bar \Psi_{k}^{(0)}\rangle$ can be chosen in
such a way that $\langle{\bf r} | \bar \Psi_{k}^{(0)}\rangle$ is a
real quantity, $\langle \bar\Psi_{k}^{(0)}|{\cal
O}|\bar\Psi_{k'}^{(0)}\rangle =-\langle \bar\Psi_{k'}^{(0)}|{\cal
O}|\bar\Psi_{k}^{(0)}\rangle$ and $\langle \bar\Psi_{k}^{(0)}|{\cal
E}|\bar\Psi_{k'}^{(0)}\rangle =\langle \bar\Psi_{k'}^{(0)}|{\cal
E}|\bar\Psi_{k}^{(0)}\rangle$.  Using these relations it follows that:
\begin{eqnarray}
\sum_{o,o'} \langle \bar \Psi_{o}^{(0)}| {\cal O}| \bar
\Psi_{o'}^{(0)}\rangle\langle \bar \Psi_{o'}^{(0)}|{\cal E}| \bar
\Psi_{o}^{(0)}\rangle&=&-\sum_{o,o'} \langle \bar \Psi_{o'}^{(0)}|
{\cal O}| \bar \Psi_{o}^{(0)}\rangle\langle \bar \Psi_{o}^{(0)}|{\cal
E}| \bar \Psi_{o'}^{(0)}\rangle \nonumber \\ &=&-\sum_{o,o'} \langle
\bar \Psi_{o}^{(0)}| {\cal O}| \bar \Psi_{o'}^{(0)}\rangle\langle \bar
\Psi_{o'}^{(0)}|{\cal E}| \bar \Psi_{o}^{(0)}\rangle,
\label{zero}
\end{eqnarray}
where for the last equality we just interchanged the dummy indexes $o$
and $o'$. From Eq. (\ref{zero}) we conclude that the double summation
of Eq. (\ref{spartial}) is equal to zero and:
\begin{equation}
{ s}=-4\sum_o{\rm Re}\left[\frac{1}{i}\langle \bar \Psi_{o}^{(0)}|
{\cal O} {\cal E}| \bar \Psi_{o}^{(0)}\rangle\right]=2\sum_o\langle
\bar \Psi_{o}^{(0)}| \frac{1}{i} [{\cal E},{\cal O}]| \bar
\Psi_{o}^{(0)}\rangle.
\end{equation}
>From this expression we finally obtain the generalized $f$-sum rule:
\begin{equation}
2\sum_o\langle \bar \Psi_{o}^{(0)}| \frac{1}{i} [{\cal E},{\cal O}]|
\bar \Psi_{o}^{(0)}\rangle=-4\sum_{o}{\rm Re}\left[\langle \bar
\Psi_{o}^{(0)}|{\cal O}{\cal G}(\varepsilon_0)\frac{1}{i}[{\cal
E},\bar H^{(0)}]| \bar \Psi_{o}^{(0)}\rangle\right].
\label{f-sumrule}
\end{equation}

\onecolumn
\newpage 
\widetext

\begin{table}
\caption{Isotropic absolute chemical shifts calculated using the IGAIM
method by Gregor \emph{et al}\cite{gregor99} and the corresponding
GIPAW-LDA results. The GIPAW calculations were performed using a
plane-wave cut-off of 100 Ry and in a 6000 Bohr$^3$ simulation
cell. 
With ``bare'', ``$\Delta$d'', and ``$\Delta$p'', we indicate the
valence GIPAW contributions to the chemical shifts, given by the
bare field ${\bf B}^{(1)}_{\rm bare}({\bf R})$ and the two correction 
fields ${\bf B}^{(1)}_{\rm \Delta d}({\bf R})$ and 
${\bf B}^{(1)}_{\rm \Delta p}({\bf R})$, respectively.
The core
contribution to the GIPAW chemical shifts is assumed to be constant
and evaluated in an all-electron atomic code.
All quantities are given as ppm.}
\label{IGAIMvsGIPAW}
\begin{tabular}{lrrrrrr}
Molecule & \multicolumn{5}{c}{$\sigma_{\rm GIPAW}$}& $\sigma_{\rm IGAIM}$ \\
 &Core&bare&$\Delta$d&$\Delta$p&Total & Total              \\\hline
H atom         &   ---                                               \\
~~CH$_4$       &   0.00 &   30.47 & 0.40 &    0.00 &  30.87 & 30.99  \\
~~CH$_3$F      &   0.00 &   25.71 & 0.41 &    0.00 &  26.13 & 26.50  \\
~~C$_6$H$_6$   &   0.00 &   22.33 & 0.41 &    0.00 &  22.74 & 23.25  \\
~~TMS          &   0.00 &   30.41 & 0.40 &    0.00 &  30.80 & 31.02  \\ 
~~SiH$_3$F     &   0.00 &   24.92 & 0.38 &    0.00 &  25.30 & 25.13  \\
~~Si$_2$H$_4$  &   0.00 &   24.53 & 0.36 &    0.00 &  24.90 & 24.78  \\
~~SiH$_4$      &   0.00 &   26.96 & 0.37 &    0.00 &  27.33 & 27.28  \\
C atom         & $1s$                                                \\
~~CO           & 198.88 & -126.25 & 4.59 & -100.15 & -22.93 & -21.16 \\
~~CH$_4$       & 198.88 &   16.86 & 3.97 &  -28.76 & 190.96 & 191.22 \\
~~CH$_3$F      & 198.88 &  -49.64 & 3.93 &  -54.70 &  98.47 &  99.66 \\
~~CH$_3$NH$_2$ & 198.88 &  -13.98 & 3.91 &  -39.05 & 149.77 & 150.44 \\
~~C$_6$H$_6$   & 198.88 &  -89.51 & 4.07 &  -77.32 &  36.12 &  39.52 \\
~~CF$_4$       & 198.88 &  -92.12 & 3.51 &  -76.05 &  34.22 &  35.29 \\
~~TMS          & 198.88 &    9.12 & 3.97 &  -32.65 & 179.33 & 182.08 \\
Si atom        &  $1s2s2p$                                           \\
~~SiF$_4$      & 832.39 &  -19.43 & 5.28 & -408.26 & 409.97 & 409.69 \\
~~SiH$_3$F     & 832.39 &  -19.50 & 5.70 & -510.30 & 308.29 & 305.45 \\
~~Si$_2$H$_4$  & 832.39 &   -9.04 & 5.80 & -622.45 & 206.70 & 202.99 \\
~~SiH$_4$      & 832.39 &   -0.21 & 5.98 & -410.20 & 427.97 & 424.37 \\
~~TMS          & 832.39 &  -17.39 & 5.70 & -518.00 & 302.70 & 304.39 \\
P atom         &  $1s2s2p$                                           \\
~~PF$_3$       & 902.47 &  -32.94 & 6.08 & -697.61 & 178.00 & 172.52 \\
~~P$_2$        & 902.47 &  -33.84 & 7.58 &-1236.95 &-360.75 &-375.45 \\
~~P$_4$        & 902.47 &   49.84 & 7.42 & -126.79 & 832.94 & 826.62 \\
\end{tabular}
\end{table}

\begin{table}
\caption{Comparison of the three different GIPAW approaches described
in Section \ref{summary}. The GIPAW-LDA calculations were performed
using a plane-wave cut-off of 100 Ry and in a 6000 Bohr$^3$ simulation
cell. The total isotropic chemical shifts are given as ppm.}
\label{IGAIMvs3xGIPAW}
\begin{tabular}{lrrr}
Molecule       & Molecular & Molecular sum rule & Crystal \\\hline
H atom                                    \\
~~CH$_4$       &  30.75 &  30.76 & 30.87  \\
~~CH$_3$F      &  26.02 &  26.01 & 26.13  \\
~~C$_6$H$_6$   &  22.69 &  22.69 & 22.74  \\
~~TMS          &  30.76 &  30.76 & 30.80  \\
~~SiH$_3$F     &  25.40 &  25.40 & 25.30  \\
~~Si$_2$H$_4$  &  24.92 &  24.93 & 24.90  \\
~~SiH$_4$      &  27.57 &  27.58 & 27.33  \\
~~
C atom                                    \\
~~CO           & -22.92 & -22.90 & -22.93 \\
~~CH$_4$       & 191.08 & 191.09 & 190.96 \\
~~CH$_3$F      &  98.53 &  98.52 &  98.47 \\
~~CH$_3$NH$_2$ & 149.61 & 149.62 & 149.77 \\
~~C$_6$H$_6$   &  36.13 &  36.14 &  36.12 \\
~~CF$_4$       &  34.62 &  34.30 &  34.22 \\
~~TMS          & 179.17 & 179.19 & 179.33 \\
Si atom                                   \\
~~SiF$_4$      & 410.12 & 409.85 & 409.97 \\
~~SiH$_3$F     & 308.27 & 308.23 & 308.29 \\
~~Si$_2$H$_4$  & 206.50 & 206.49 & 206.70 \\
~~SiH$_4$      & 427.95 & 427.95 & 427.97 \\
~~TMS          & 302.61 & 302.61 & 302.70 \\
P atom                                    \\
~~PF$_3$       & 177.90 & 177.70 & 178.00 \\
~~P$_2$        &-360.97 &-360.97 &-360.75 \\
~~P$_4$        & 832.87 & 832.87 & 832.94 \\
\end{tabular}
\end{table}

\begin{table}
\caption{The NMR chemical shift for carbon in methane using
Troullier-Martins potentials with a range of core radii. These LDA
calculations were performed using a plane-wave cut-off of 180 Ry
(converged to 0.01 ppm for the hardest potential) and in a simulation
cell of 1000 Bohr$^3$.
With ``bare'', ``$\Delta$d'', and ``$\Delta$p'', we indicate the
valence GIPAW contributions to the chemical shifts, given by the
bare field ${\bf B}^{(1)}_{\rm bare}({\bf R})$ and the two correction 
fields ${\bf B}^{(1)}_{\rm \Delta d}({\bf R})$ and 
${\bf B}^{(1)}_{\rm \Delta p}({\bf R})$, respectively.}
\label{psp-methane}
\begin{tabular}{lrrrrr}
Core radius (Bohr) & \multicolumn{5}{c}{$\sigma_{\rm GIPAW}$} \\
     &  Core  & Bare  &$\Delta$d &$\Delta$p & Total           \\\hline
1.2  & 198.88 &  7.30 &    3.96  &  -19.14  & 191.00          \\
1.4  & 198.88 & 12.22 &    3.99  &  -24.08  & 191.01          \\
1.6  & 198.88 & 17.03 &    3.98  &  -28.64  & 191.25          \\
1.8  & 198.88 & 21.65 &    3.92  &  -32.86  & 191.59          \\
\end{tabular}
\end{table}

\begin{table}
\caption{The valence contribution to the isotropic chemical shift of crystalline diamond (ppm).}
\label{sigma-diamond}
\begin{tabular}{lc}
Method & Valence contribution to $\sigma$\\\hline
GIPAW                    & -65.85        \\
All-electron at 800 Ry   & -64.89        \\
All-electron at 1400 Ry  & -64.46 
\end{tabular}
\end{table}

\end{document}